\documentclass{osa-article}
\usepackage{subcaption} 
\usepackage{color,soul}
\usepackage{hyperref}
\journal{oe}

\articletype{Research Article}

\usepackage{lineno}

\begin{document}

\title{Hyperpixels: Pixel Filter Arrays of Multivariate Optical Elements for Optimized Spectral Imaging}

\author{Calum Williams \authormark{1}, Richard Cousins \authormark{2}, Christopher J. Mellor \authormark{3}, Sarah E. Bohndiek,\authormark{4,5} and George S.D. Gordon,\authormark{6,*}}

\address{\authormark{1}Department of Physics, University of Exeter, Exeter, EX4 4QL, United Kingdom}
\address{\authormark{2}Nano- and Micro-scale Research Centre, University of Nottingham, UK, University Park, Nottingham NG7 2RD, UK}
\address{\authormark{3}School of Physics and Astronomy, University of Nottingham, University Park, Nottingham NG7 2RD, UK}
\address{\authormark{4}Department of Physics, Cavendish Laboratory, University of Cambridge, JJ Thomson Avenue,Cambridge, CB3 0HE, UK}
\address{\authormark{5}, Cancer Research UK Cambridge Institute, University of Cambridge, Robinson Way, Cambridge, CB2 0RE, UK}
\address{\authormark{6} Optics \& Photonics Research Group, Department of Electrical and Electronic Engineering, University of Nottingham, Nottingham, NG7 2RD, UK}

\email{\authormark{*}george.gordon@nottingham.ac.uk} 



\begin{abstract}
We introduce the concept of `hyperpixels' in which each element of a pixel filter array (suitable for CMOS image sensor integration) has a spectral transmission tailored to a target spectral component expected in application-specific scenes. These are analogous to arrays of multivariate optical elements that could be used for sensing specific analytes. Spectral tailoring is achieved by engineering the heights of multiple sub-pixel Fabry-Perot resonators that cover each pixel area. We first present a design approach for hyperpixels, based on a matched filter concept and, as an exemplar, design a set of 4 hyperpixels tailored to optimally discriminate between 4 spectral reflectance targets. Next, we fabricate repeating 2$\times$2 pixel filter arrays of these designs, alongside repeating 2$\times$2 arrays of an optimal bandpass filters, perform both spectral and imaging characterization. Experimentally measured hyperpixel transmission spectra show a 2.4x reduction in unmixing matrix condition number ($p$=0.031) compared to the optimal band-pass set. Imaging experiments using the filter arrays with a monochrome sensor achieve a 3.47x reduction in unmixing matrix condition number ($p$=0.020) compared to the optimal band-pass set. This demonstrates the utility of the hyperpixel approach and shows its superiority even over the optimal bandpass case.  We expect that with further improvements in design and fabrication processes increased performance may be obtained. Because the hyperpixels are straightforward to customize, fabricate and can be placed atop monochrome sensors, this approach is highly versatile and could be adapted to a wide range of real-time imaging applications which are limited by low SNR including micro-endoscopy, capsule endoscopy, industrial inspection and machine vision.
\end{abstract}

\section{Introduction}
Spectral imaging is a powerful technique that captures wavelength-resolved information (spectra) at every pixel in an image to non-invasively estimate relative abundances of spectral signatures, representing for example the chemical composition of spatially heterogeneous samples  \cite{liReviewSpectralImaging2013, Lu2014, spigulisSingleSnapshotMultispectral2012, bauerSpectralFilterArray2019}. The interaction of light with a given sample, for example through absorption, reflectance or fluorescence, provides a spectral signature that is determined by the relative abundance of key chromophores or fluorophores. Applying `spectral unmixing' algorithms to extract chemical abundances can then provide valuable diagnostic information across a wide range of disciplines, from biomedical imaging to agricultural inspection \cite{Waterhouse2019, clancySurgicalSpectralImaging2020, siriluthmanBimodalReflectanceFluorescence2019,liuHyperspectralImagingTechnique2017, clarkeUsingHyperspectralImagery2021, adaoHyperspectralImagingReview2017}.

Real-time spectral imaging is commonly realized using multispectral filter arrays (MSFAs). MSFAs are mosaics of pixelated optical filters integrated atop an image sensor; images at each spectral band are formed by interpolation between spatially distributed filters using a demosaicking algorithm \cite{liReviewSpectralImaging2013, Sawyer2022,wuOptimizedMultiSpectralFilter2019,williamsGrayscaletocolorSinglestepFabrication2019}. Conventionally, each element in the MSFA is an optical bandpass filter, usually with a fixed, narrow bandwidth. The center wavelengths of each spectral band can either be spaced evenly across the spectral range of the sensor (e.g. 400 -- 900 nm for silicon image sensors) for generic applications, or can be spectrally and spatially optimized to a particular application via computational processes \cite{bauerSpectralFilterArray2019,wuOptimizedMultiSpectralFilter2019,Sawyer2022}. One key shortcoming of bandpass filters is that much of the incoming spectral power---outside of the bandpass region---is discarded (reflected or absorbed). This can reduce performance, in particular the signal-to-noise ratio (SNR), which may be problematic in environments with high noise or interference from other spectral signatures.

If \emph{a priori} information of the sample is known then an optimal spectral imaging approach---with regards to SNR---would be to match the spectral transmission of the filters to the spectrum of each contributing chromophore; the concept of `matched filters' is widely used for SNR maximization in telecommunications \cite{turinIntroductionMatchedFilters1960} and other fields. For optical filters, matching has been achieved using optimized thin-film multilayered elements called multivariate optical elements \cite{Cianci2000} in challenging spectroscopy applications where highly sensitive detectors cannot be easily deployed e.g. in undersea oil and gas extraction \cite{Nelson1998,Kozhukh2012}.  However, multivariate optical elements have only been used to make spatially uniform filters and have not been applied as pixelated filter arrays. 

Demonstrating a matched filter concept in an MSFA has been elusive to-date due to the complexity of creating arbitrary (i.e. not single narrow-bandpass) spectral characteristics within pixel filter arrays. Previous studies have achieved some compact, broadband solutions using: nanohole metasurfaces \cite{Xiong2022,ahamedReconstructionbasedSpectroscopyUsing}; multilayer structures with spatially varying holes etched into them \cite{Mirotznik2009}; metasurfaces designed to perform `pixel routing' where sub-wavelength 3D structures route light to different parts of the image sensor based on their wavelength \cite{Zou2022}; random or semi-randomized diffusers \cite{Monakhova2020}; and coded apertures \cite{wagadarikarSingleDisperserDesign2008}. Nonetheless, it is challenging to tailor spectra with these approaches due to  computationally intensive inverse design processes and designs typically require high resolution (nanoscale) lithographic techniques with tight tolerances. Most recently MSFAs with each pixel made up of a Fabry-Perot etalon with random cavity height (and hence resonant wavelength) have been demonstrated \cite{yakoVideorateHyperspectralCamera2023}, which also show potential for multiband transmission but with potential for an easier design and fabrication process.

\begin{figure}[!htpb]
    \centering
    \includegraphics[width=1.0\textwidth]{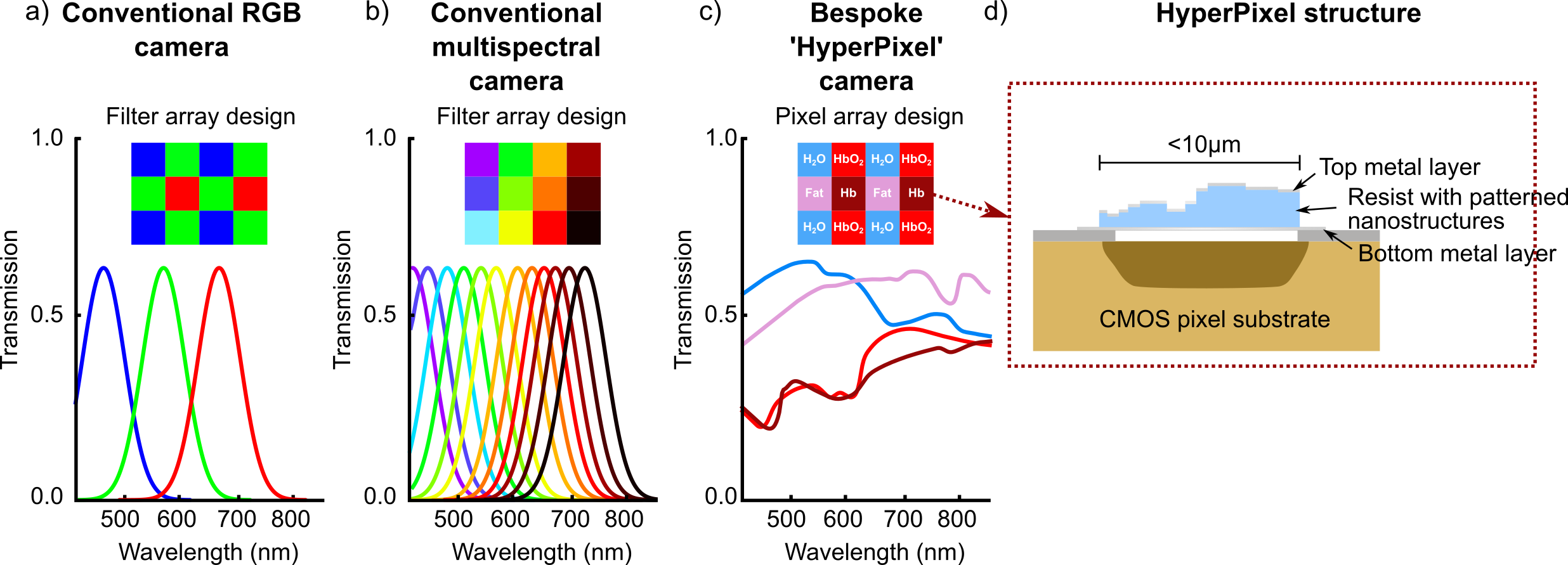}
    \caption{Hyperpixel concept: a) Conventional color cameras have Bayer filters applied per-pixel to discern red, green and blue wavelengths. b) Conventional snapshot multispectral cameras based on multispectral filter arrays (MSFAs) extend this using bandpass filters centered at many different wavelengths -- 16 in this example. c) hyperpixels replace the conventional MSFA filters and enable different pixels spectral responses to be `matched' to the reflectance spectrum of a target substance. d) Hyperpixels are implemented by fabricating several Fabry-Perot resonators within the area of a single pixel on a CMOS sensor. The area of each resonator dictates the power transmitted in that particular band, but the total power integrated across the spectrum cannot exceed that of an equivalent single band-pass filter. }
    \label{fig:concept}
\end{figure}

Building on these previous works, we demonstrate here a new matched filter concept, `hyperpixel' filter arrays (see Figure \ref{fig:concept}), that can be deployed as an MSFA for snapshot multispectral imaging. Hyperpixels comprise multiple spatially heterogeneous Fabry-Perot resonators, termed \emph{subpixel filters}, located within the bounds of a single image sensor pixel. Each of the subpixel filters can have a different cavity height, so that when integrated over the pixel area, the overall transmission is a composite spectral filter based on the sum of these responses. The hyperpixel approach overcomes the design and fabrication demands of prior approaches, being easily designed based on cavity heights and fabricated using grayscale lithography \cite{Williams2019}. As a result of the hyperpixels implementation, spectra can be easily unmixed using only a single standard matrix multiplication that can be implemented in real-time using existing cost-effective hardware. We demonstrate here the hyperpixels concept using 100 subpixel filters to tailor spectra in a 10$\times$10 $\mu$m hyperpixel targeted to four distinct reflectance spectra and arranged in a hyperpixel filter array.  We demonstrate the potential performance improvement afforded using a MacBeth ColorChecker chart compared to a standard band-pass filter array, representing the best possible conventional MSFA.

\section{Methods}
\subsection{Hyperpixel design}
\begin{figure}[!htpb]
    \centering
    \includegraphics[width=1.0\textwidth]{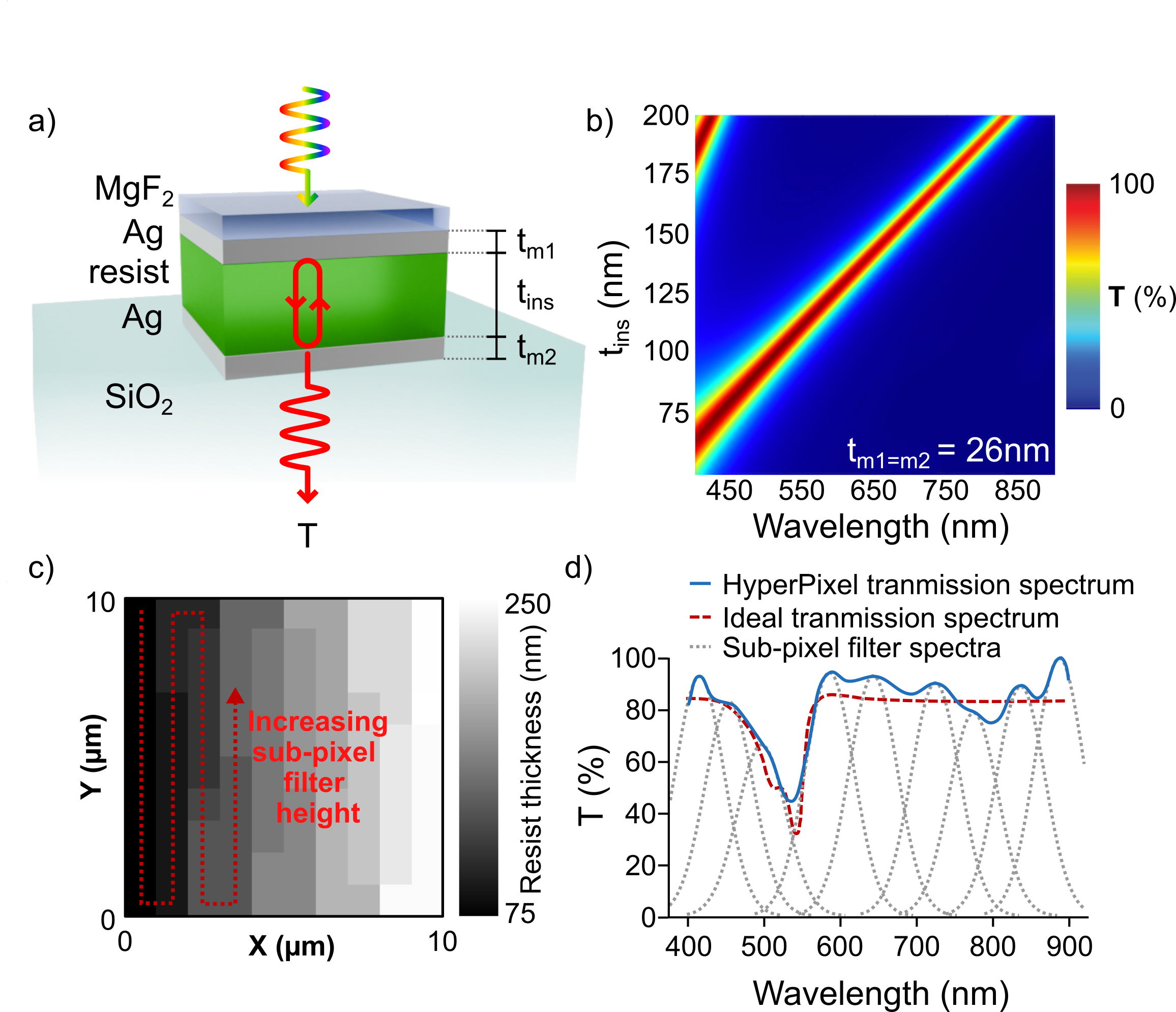}
    \caption{Design process for hyperpixels. a) Structure of constituent Fabry-Perot cavities making up each hyperpixel. $t_{m1}$ = 22nm, $t_{m2}$ = 22nm and $t_{ins}$=75--200nm. b) Simulation data showing the theoretical relationship between cavity thickness and centre transmission wavelength. c) Example spatial design of hyperpixel showing ordering by increasing height of sub-pixels for more reliable fabrication. d) Example transmission spectrum produced by hyperpixel design showing a close match to target spectrum.}
    \label{fig:hyperpixeldesign}
\end{figure}

Hyperpixels were implemented as an array of thin-film metal-insulator-metal Fabry-Perot filters bounded within the pixel active area of a CMOS image sensor (Figure \ref{fig:hyperpixeldesign}a). For alignment simplicity, we assume a 10 $\mu$m pixel size and divide this up into 100 `subpixel filter' units. The number of subpixel filter units is adjustable but should be a square number to match the geometry of the CMOS image sensor pixels.  Subpixel filters should also be $>$ 500nm to avoid performance drop-off experience as the dimension approaches the wavelength of light \cite{Goossens2021} and to minimize fabrication artefacts as structures become smaller. Working within these constraints we choose 100 subpixel filters as as a compromise between control over spectral shape and quality of subpixel filter.

Each of these subpixel filters comprises a single cavity filter with Ag mirrors (thickness 22 nm) and resist cavity (variable thickness $\sim$75-250 nm). A MgF$_2$ encapsulation (capping) layer was used to mitigate oxidization and surface tarnishing. Silica glass (hundreds of micron thickness) was used as the substrate, which permits straightforward integration atop commercial monochrome image sensors or optical alignment (using relay optics) to the image sensor. To determine the optimal multilayer thicknesses for high transmission and sufficiently narrow passbands, numerical simulations were performed using Lumerical FDTD Solutions (Ansys) (Figure \ref{fig:hyperpixeldesign}b).  

Following simulations, an initial array with evenly spaced resist thicknesses from $\sim$50 to 250nm was fabricated as a height calibration target. Actual heights were measured using a combination of atomic force microscopy (AFM) and optical micro-spectrophotometery, as described in our previous works \cite{taylor-williamsSpectrallyTailoredHyperpixel2022} and used to optimize exposure dose by calibrating the relationship between actual (measured) height and designed height. Height calibration was iterated several times until heights are matched, because the optimal exposure dose profile is dependent on numerous factors including the substrate material/s, processing parameters (i.e. development times, temperatures, proximity effect etc). Optical micro-spectrophotometery was performed on the sub-pixels to characterize their optical responses (transmission spectra) and locate relative positions of center wavelengths, second order resonances, bandwidths (full width at half maximum, FWHM) and transmission. 
 
To create a hyperpixel from the subpixel filters, we must next determine the required heights of each of the 100 subpixel filters so that they match the desired spectral transmission response.  To do this, we begin by identifying the target transmission spectrum required (e.g. Figure \ref{fig:hyperpixeldesign}d).  We must then approximate this as a sum of bandpass filters, which represent the subpixel filters.  The number of independent bandpass filters used need not be the same as the number of subpixel filters, and can be thought of as a method of quantizing the heights of the subpixel filters.  This quantization process may be important for future manufacturing. For example if one exposure is required per grayscale height level (e.g. with a stepper) fewer heights would allow faster manufacturing. Similarly, there may be tolerances on the grayscale heights meaning that only finite number of distinct heights are realistically manufacturable.  We therefore choose a quantization level of 32, which enables up to 32 distinct heights.  We find this number enables fast and reliable convergence of our design optimization algorithm with high fidelity curve approximation, while still producing good quality fabrication results.  The optimization process may assign certain certain wavelengths weights of zero, or may assign a certain centre wavelength multiple times so the final quantization level may even be less than 32.

The optimization algorithm uses experimentally fitted spectral data to map a resist thickness to a spectral transmission profile and create 32 bandpass spectra that are then summed together to get the total transmission spectral profile. The center wavelengths and weights in the final summation are set as the adjustable parameters that are then optimized using a basin-hopping algorithm. The error function is a mean-square error with the target spectrum, which is the expected reflectance spectrum of the target, including the effect of non-flat spectral profile of the input illumination. The output of this process is a composite spectrum that closely matches the target (Figure \ref{fig:hyperpixeldesign}d).  The optimization algorithm is implemented using custom Python code (available at \cite{datadoi}).

Finally, these 32 component spectra must be mapped onto the 100 available subpixel filters within each hyperpixel. To do this, centre wavelengths are first rounded to the nearest 3nm reflecting again the achievable height precision measured using AFM previously \cite{taylor-williamsSpectrallyTailoredHyperpixel2022}. The algorithm produces weights for each of these components that vary from 0 to 1.  In the final design these weights will be implemented by using the same height value for a greater number of the 100 available subpixels: greater weight for a particular component spectrum means more subpixels will have the height that creates that component.  To achieve this, the weights within each hyperpixel are normalized and rounded to the nearest integer such that they sum to 100.  These new normalized weights are then used to determine how many of the 100 subpixel filters (i.e. 10$\times$10 array) should have that particular height.

Once the subpixel filters heights are determined, they are placed in increasing height order in a snaking pattern to minimize height steps between adjacent subpixel filters, which should enable more reliable fabrication (Figure \ref{fig:hyperpixeldesign}c).

\subsection{Filter array design}
\begin{figure}[!htpb]
    \centering
    \includegraphics[width=1.0\textwidth]{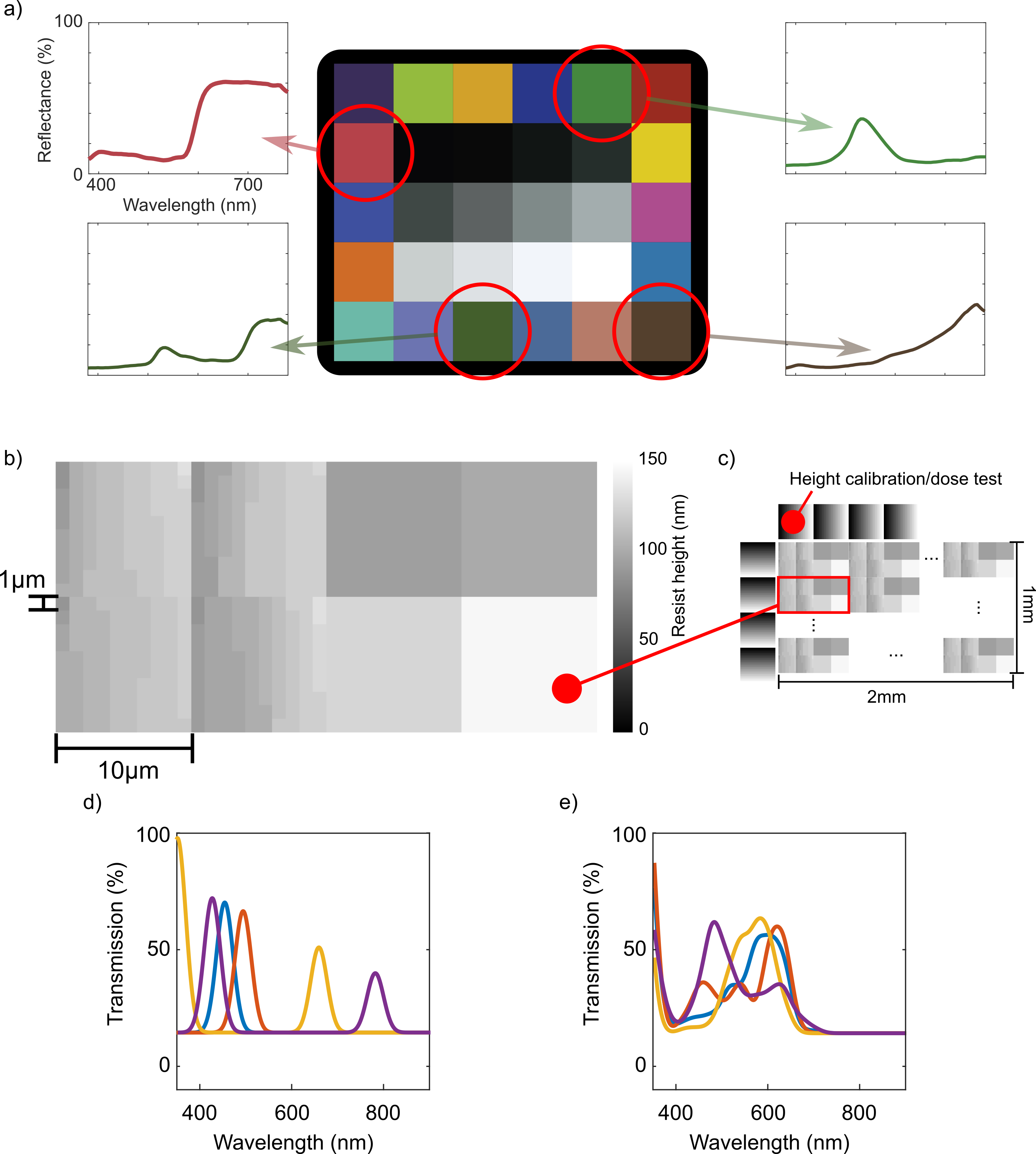}
    \caption{Designing hyperpixels for optimal spectral unmixing. a) Selected spectral reflectance targets on a Macbeth color checker chart, showing spectral reflectance profiles. b) Unit cell of designed hyperpixel filter array, including 8 10$\times$10 $\mu$m pixel designs: 4 hyperpixel designs (comprising 10$\times$10 1$\mu$ subpixel filters) and 4 bandpass filters with optimized centre wavelengths. c) Layout of the full filter array as fabricated, including height calibration targets around the edge. f) Simulated spectra of bandpass filter pixels from filter array. e) Simulated spectra of hyperpixel filters from filter array.}
    \label{fig:macbethdesign}
\end{figure}

To design a test set of filter arrays to evaluate performance of hyperpixels against conventional bandpass filters, we identified a subset of target spectra from a MacBeth ColorChecker chart (Figure \ref{fig:macbethdesign}a) and designed hyperpixels optimally matched to them.  We combined manufacturer provided spectral reflectance data with the spectrum of the illumination source to be used (OSL2 Halogen source, ThorLabs) to determine the total expected reflected spectral profile. These spectra were used as the target design spectra for the hyperpixels.

Four hyperpixel designs were produced as shown in Table \ref{tab:HP_designs}.

\begin{table}[]
\centering
\caption{Specifications of designed set of four hyperpixel filters.}
\label{tab:HP_designs}
\begin{tabular}{|c|c|c|c|c|}
\hline
~  & \textbf{HP filter 1} & \textbf{HP filter 2} & \textbf{HP filter 3} & \textbf{HP filter 4} \\ \hline
Height (nm) & \multicolumn{4}{|c|}{No. subpixel filters} \\ \hline
\textbf{87}  & 2                            & 3                            & 1                            & 2                            \\ \hline
\textbf{93}  & 3                            & 10                           & 1                            & 8                            \\ \hline
\textbf{96}  & 4                            & 8                            & 0                            & 26                           \\ \hline
\textbf{99}  & 0                            & 0                            & 2                            & 0                            \\ \hline
\textbf{102} & 11                           & 6                            & 12                           & 16                           \\ \hline
\textbf{108} & 8                            & 13                           & 23                           & 8                            \\ \hline
\textbf{111} & 24                           & 3                            & 0                            & 9                            \\ \hline
\textbf{114} & 0                            & 0                            & 27                           & 0                            \\ \hline
\textbf{117} & 24                           & 27                           & 25                           & 10                           \\ \hline
\textbf{123} & 22                           & 29                           & 9                            & 14                           \\ \hline
\textbf{126} & 1                            & 1                            & 0                            & 5                            \\ \hline
\textbf{132} & 1                            & 0                            & 0                            & 2                            \\ \hline
\end{tabular}
\end{table}

As a benchmark we also designed 4 bandpass filters optimized for spectral unmixing of the same 4 selected targets.  Specifically, for the set of 4 bandpass filters, parameterized by 4 centre wavelengths, we multiply the filter spectral transmission profile with the expected spectral reflectance of each of the target areas. The output is the $4\times4$ matrix which, when inverted, produces an `unmixing' matrix that recovers the correct abundances of the target spectra.  The condition number of this matrix represents how `well' the unmixing can be performed in the presence of noise and so we minimize this quantity as our optimization objective to produce optimal performance of the bandpass filters \cite{gaoOpticalHyperspectralImaging2015}. Similarly, the condition number of the hyperpixel filter set can also be computed. The ratio of these condition numbers is then computed to determine whether the hyperpixels should perform better or worse than the optimal bandpass filters for a particular set of spectral targets:

\begin{equation}
r = \mathrm{cond}(\mathbf{H}_{bandpass}) / \mathrm{cond}(\mathbf{H}_{hyperpixel})
\end{equation}

\noindent where $\mathbf{H}_{bandpass}$ is the $4 \times 4$ unmixing matrix of the optimized bandpass filters and $\mathbf{H}_{hyperpixel}$ is the unmixing matrix for the hyperpixel filters. For the set of spectral reflectance targets in the MacBeth color chart, we find that $r$ can vary from 0.1 (bandpass filters perform better) up to to 1.6 (hyperpixels perform better).

Finally, we create a layout for fabrication of the filter arrays with the 4 hyperpixels adjacent to the bandpass filters (Figure \ref{fig:macbethdesign}b). By placing these regions together in a repeating array, we ensure that any localized fabrication variations (e.g. proximity effect from e-beam lithography, resist thickness etc) apply equally to all the different designs. The unit cells have dimension 20$\mu$m$\times$20$\mu$m and are repeated into an array of 1mm $\times$ 2mm (Figure \ref{fig:macbethdesign}c).  Around the edge of the array are height calibration targets with a linear ramp of 0 to 250nm so that fabricated heights be validated with AFM or stylus profiler.

Using our previously described calibration method, which combines AFM and with optical micro-spectrophotometery and simulations (Figure \ref{fig:hyperpixeldesign}b), we map the target heights to expected spectral profiles for the bandpass filters, as shown in Figure \ref{fig:macbethdesign}d.  Similarly, we can also produce expected spectra for the hyperpixels by including second-order resonances using our previously described characterization approach \cite{taylor-williamsSpectrallyTailoredHyperpixel2022}.

\subsection{Fabrication}
22 mm high precision glass coverslips (ThorLabs) were used as the substrates. Before processing, these were successively cleaned in ethyl lactate, acetone, methanol, and isopropanol (IPA), then dehydrated on a contact hotplate at 150$^o$C for 3 minutes. For the first mirror layer, a thermal evaporator was used to deposit: a 1 nm Ti adhesion layer followed by 22 nm Ag and MgF$_2$ capping layer. The Ag was evaporated at a high deposition rate to improve the film quality.

As soon as possible after evaporation, a film of AR-7720 (Allresist GmBh) was spin-coated at 4000 rpm for 60 s. The resist was diluted using PGMEA (1-methoxy-2-propanol acetate), so an initial film thickness of 165 nm was produced. The sample was then placed in an electron beam lithography tool, NanoBeam nB5 (NanoBeam Ltd.). To vary the height of each pixel, the dose was varied using an earlier dose test which gave the contrast as 0.55. The exposed sample then had a post-exposure bake on a hotplate at 110$^0$C for 60 s followed by a bake in an oven at 90$^o$C for 2 hours; this latter step reduces the surface roughness of the film. The resist was developed in a TMAH (Tetramethylammonium hydroxide) based developer for 90 s and then rinsed with DI (deionized) water. This resulted in a resist pattern with varying heights. The second mirror layer was thermally evaporated,  consisting of 22 nm of Ag and MgF$_2$ encapsulation layer to prevent silver tarnishing, and is kept to $<$ 30 nm. 

\subsection{Spectroscopic characterization}
For spectral characterization, a modified Olympus BX-51 optical microscope was used to analyse the hyperpixel filter arrays. A UV-visible spectrometer (Ocean Optics HR2000+) was used to collect the spectral transmission data, while a digital camera (Lumenera Infinity-2 2MP CCD) was used to image the surface. The spectra collected were normalized using OceanView (Ocean Optics) software by collecting the bright and dark reference spectra. The bright spectra is obtained by using a non-metal coated area of the filter array i.e. just the light source transmitting through the glass.

The spectra of each sub-pixel was collected using a series 100x objective and centering the collection area over the center of the pixel. 4 spectra (at different spatial positions in the adjacent array areas) were collected and averaged to reduce noise. 

In order to examine the variations in fabrication quality across the sample, 14 different regions across the fabricated filter array were sampled and spectra recorded from each bandpass and hyperpixel in the unit cell.  The spectra from the bandpass filters were further analysed by fitting a sum of two Gaussian curves to identify the first- and second-order Fabry-Perot resonance wavelengths and the FWHM. Direct measurement of the hyperpixel spectra is challenging because the circular aperture of the spectrometer collection fibre is mismatched with the square aperture of the pixel. Therefore, the bandpass filter spectra are used as a measure of deviation from designed height in each of the 14 sampled areas. 

First, the centre wavelength of each of the four bandpass areas is measured and plotted against the target height from the original design.  A line is fitted to this.  Then the target heights of the each of the subpixel filters in the hyperpixel filters are input into this fitted line and an adjusted actual centre wavelength for each of the subpixel filters is found.  This adjusted centre wavelength, along with other parameters from the BP areas such as the FWHM and position of second-order resonance, are used to construct a transmission for each subpixel area.  These are then summed to infer the overall hyperpixel filter response in this area of the filter array.

Based on these measured bandpass and inferred hyperpixel spectra, an unmixing matrix is formed by multiplying these transmission spectra by the ideal reflected spectra expected from the target areas on the color chart. The condition numbers of these matrices for the hyperpixel and bandpass areas are then computed and compared using a paired $t$-test.

\subsection{Spectral imaging characterization}
Our validation experiment was to image a real Macbeth color chart target using the fabricated filter arrays placed on top of a CMOS sensor image sensor. An optical imaging relay system was constructed for spectral imaging. The system contains a first imaging lens: fixed-focal length camera lens (f=35mm, Navitar), and a set of relay lenses (Thorlabs, AC254-075-A-ML). The target scene, illuminated by a halogen light source coupled to a ring light illuminator (Thorlabs OSL2IR), is imaged onto a monochrome image sensor (Basler, a2A3840-45umPRO) via a conjugate plane located between two 4x planar achromat objectives (Thorlabs, RMS4X). The conjugate plane contains the hyperpixels mounted to a XYZ-translation stage (Thorlabs, MBT616D). Imaging the scene through the filter pixels in this way mimics a conventional snapshot multispectral camera whereby MSFAs are integrated directly atop the pixels.

\subsection{Image processing}

Raw monochrome images from the camera were first straightened and cropped such that only the color chart area was included. Images were then demosaicked to produce multispectral image data cubes by manually identifying the same pixel type in 5-10 separate locations across the image then fitting a grid to these locations to determine the periodicity, location of each pixel type and offset (custom interactive MATLAB code, available from \cite{datadoi}). Once this grid is determined, the raw pixels within each filter array pixel are summed to form a single point in the hypercube. The demosaicking and interpolation process represents a reduction in resolution from the raw images: the raw images are $3840 \times 2160$ but each pixel in the pixel filter array covers around 11.05 sensor pixels as the pixel sizes were not matched to the sensors.  The raw demosaicked images were therefore $174 \times 49$ pixels. Because the image height is scaled down twice as much due to the rectangular shape of the unit cell, the images in the hypercube are resampled to $1920 \times 1080$ to preserve aspect ratio.

A second processing step was then performed on the demosaicked images to identify each of the different spectral reflectance areas within the color chart. First, the flat spectral reflectance areas (white and black) were identified and used in a similar fashion to white and dark references in spectroscopy; pixels within these areas are averaged to create white and dark reference, $r_{white}$ and $r_{dark}$ respectively, for each of the channels in the hypercube. The dark reference was subtracted from each of the channels in the hypercube and then the white reference was used as a normalization value to divide values in each channel.  Specifically:
\begin{equation}
\hat{p}(x,y) = \frac{p(x,y) - r_{dark}}{r_{white} - r_{dark}}
\end{equation}
\noindent where $\hat{p}(x,y)$ is the corrected pixel value at position $(x,y)$, $p(x,y)$ is the raw pixel value and $r_{white}$ and $r_{dark}$ are the white and dark reference values respectively.

Because of the fixed power constraint of the hyperpixel design, we expect that all pixels in the white reference areas should transmit approximately the same total power and so this normalization is applied independently to each channel in the BP and hyperpixel filter arrays.

Finally, the four target spectral areas are identified. The pixel values within each of these 4 areas are averaged and this is repeated for the 4 channels representing the hyperpixels and bandpass filters. The result is 16 values for each of the bandpass and hyperpixel cases, which are arranged into the $4 \times 4$ unmixing matrices, similarly to the spectral analysis case. The condition numbers of these matrices represent quality of unmixing. For visual comparison, the 4 target regions in the hypercube images are unmixed, using the computed spectral unmixing matrix, and are plotted in a $2 \times 2$ array for each of the hyperpixel and bandpass cases. To quantify the effect of fabrication variation across the sample on performance, the unmixing matrices are computed for 10 different areas across the pixel filter array and a $t$-test is used to establish whether a statistically significant reduction in condition number is achieved.

\section{Results}
\subsection{Pixel-wise spectroscopic characterization}
\begin{figure}[!htpb]
    \centering
    \includegraphics[width=1.0\textwidth]{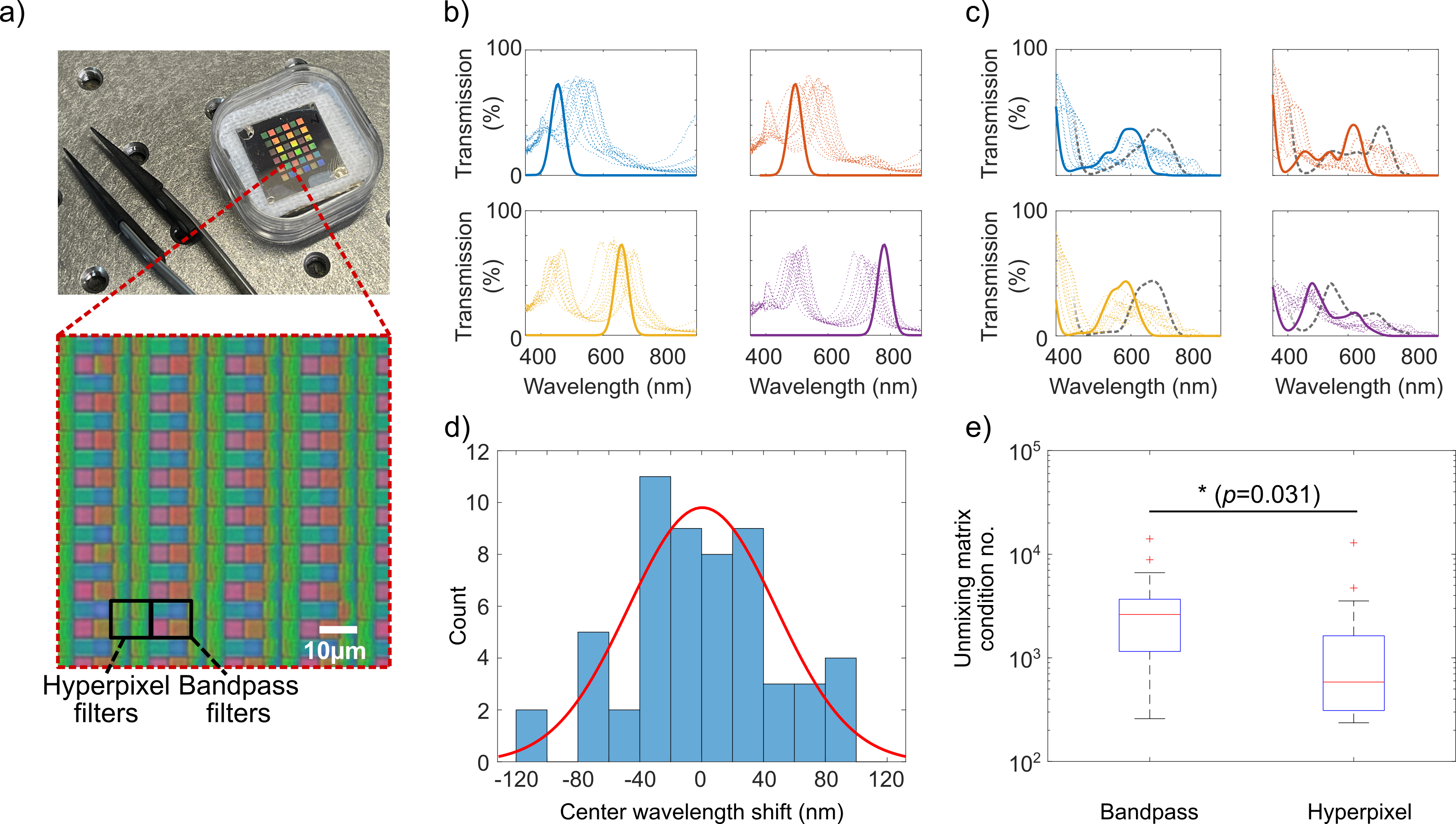}
    \caption{Spectral characterization of fabricated hyperpixel filter array. a) Microscope image showing repeating $2 \times 4$ array of hyperpixels and bandpass filters. b) Measured spectra from bandpass filters. Solid lines: designed spectra, dotted lines: measured spectra from 14 different areas on filter array. c) Reconstructed spectra from hyperpixel filters. Dashed lines: designed spectra, solid lines: blue-shifted spectra following fabrication, dotted lines: inferred spectra from 14 different areas on filter array. d) Histogram showing error between designed centre wavelength and measured centre wavelength for bandpass filters across 14 locations on filter array (normal distribution fit: $\mu$=0, $\sigma$=47.4nm. e) Comparison of condition number of unmixing matrix shows a statistically significant reduction (ratio = 2.41x, 95\% CI = 1.10 -- 5.30) of the hyperpixels compared to optimal bandpass filters.}
    \label{fig:spectraresults}
\end{figure}

Experimentally fabricated filter arrays (Figure \ref{fig:spectraresults}a) were used to obtain spectral response (transmission) results for 14 different spatial regions (Figure \ref{fig:spectraresults}) from the centre of the array to the outer edges.  The measured transmission spectra from the bandpass filters (Figure \ref{fig:spectraresults}b) indicate a relatively large variation in central peak position across the sample, though some areas do closely match the designed structures. The inferred spectra for the hyperpixels (Figure \ref{fig:spectraresults}c. were reconstructed from the corresponding wavelength shift of adjacent bandpass filters. By aggregating the observed shifts in centre wavelength of the bandpass filters (Figure \ref{fig:spectraresults}d) a normal distribution fit can be made, which gives a standard deviation of 47.7 nm in centre wavelength across our sample.

Due to a fabrication specification error, the hyperpixel spectral are blue-shifted by 50-100nm (dashed lines in Figure \ref{fig:spectraresults}c) compared to the target designs (solid lines in Figure \ref{fig:spectraresults}c).  This may reduce performance, however a close inspection of the spectra measured from hyperpixels in different filter array regions (dotted lines in Figure \ref{fig:spectraresults}c) shows that the variation in spectral shift across the filter array is comparable in magnitude to this blue shift and in many cases actually compensates.  At a minimum we therefore expect that at least some areas of the sample will exhibit hyperpixel spectra matching the design specification. Further, we note that since this fabrication blueshift did not occur with the bandpass filters these ought to closely match the optimal designs (save for variation across the sample). Any observed performance advantage of hyperpixels is therefore still being measured against an optimal benchmark. Such variations can be minimized with tighter control over fabrication tolerances and processes in future.

Despite the limitations in the fabrication process, which led to inaccuracies in the filter properties compared to design, the measured condition number for the unmixing matrices of bandpass and hyperpixel filters shows that hyperpixels produce statistically significantly lower condition numbers than optimized bandpass filters (ratio = 2.41x, 95\% CI = 1.10 -- 5.30, $p$=0.031) for this set of reflectance targets (Figure \ref{fig:spectraresults}e).

\subsection{Spectral imaging characterization}
\begin{figure}[!htpb]
    \centering
    \includegraphics[width=1.0\textwidth]{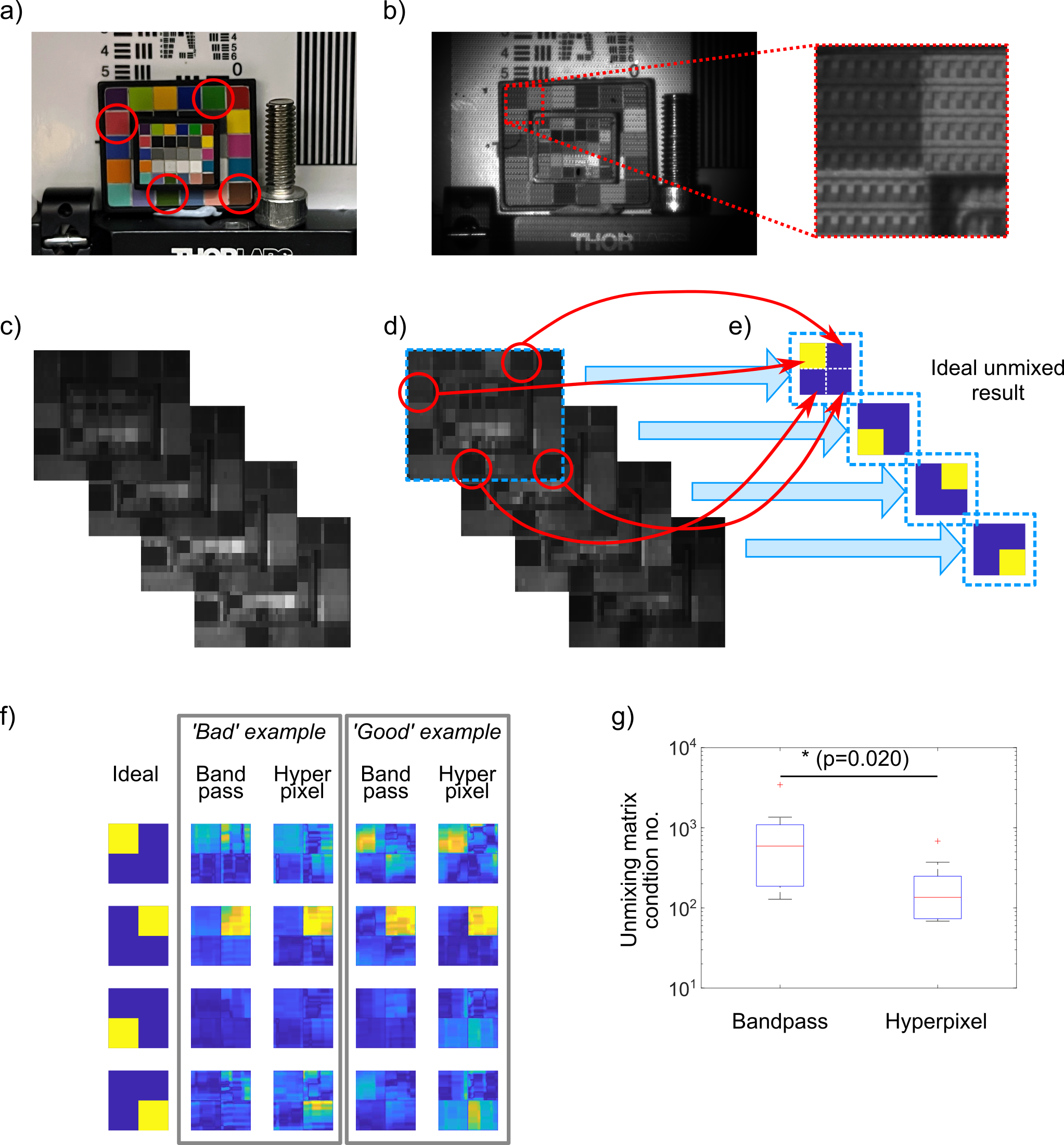}
    \caption{Imaging experiments with hyperpixel filter array. a) Conventional RGB image of sample scene.  b) Raw image recorded with filter array placed atop CMOS sensor, inset shows the mosaic pattern. c) Example demosaicked hypercube of images for bandpass filters. d) Example demosaicked hypercube of images for hyperpixel filters. e) The appropriate spectral reflectance target areas are cropped from the full image frames, and then are spectrally unmixed using the appropriate 4$\times$4 unmixing matrix to produce 4 images.  Under ideal conditions of perfect unmixing, only one of the 4 spectral reflectance targets should contain power in any recovered frame. f) Unmixed images for a section of the filter array where the hyperpixel outperforms the optimal band-pass filter in terms of condition number, and also a section where hyperpixel performance is lower than optimal bandpass.  g) Over 10 different regions of the filter array we find that the condition number of the unmixing matrix is significantly reduced for hyperpixels compared to bandpass (ratio = 3.47x, 95\% CI =  1.28 -- 9.40, $p$ = 0.020).}
    \label{fig:imagingresults}
\end{figure}

Spectral imaging data were next obtained using the MacBeth Color Checker chart scene (Figure \ref{fig:imagingresults}a, obtained using a smartphone camera; target reflectance squares are circled).  The raw unprocessed image through the hyperpixel filter array (Figure \ref{fig:imagingresults}b; mosaic visible on zoomed inset) was demosaiced to produce 2 spectral hypercubes: bandpass (Figure \ref{fig:imagingresults}c) and hyperpixel (Figure \ref{fig:imagingresults}d).  Each hypercube comprises 4 planes representing the different spectral channels.  From these hypercubes, relevant sections of the image were cropped to highlight the target spectral reflectance areas (Figure \ref{fig:imagingresults}d and e).

Due to variations in fabrication quality across the filter array, we find that relative benefit of the hyperpixels compared to bandpass filters varies. We present an example with greater reduction in condition number (Figure \ref{fig:imagingresults}f).  Though reconstruction is still far from ideal, we observe that for hyperpixels, more power is directed to the appropriate pixel than for the bandpass filters, in which the power is largely directly to the first 2 pixels only (particularly in the 3rd and 4th columns of Figure \ref{fig:imagingresults}f).  This visible benefit reduces somewhat in a `bad' example (Figure \ref{fig:imagingresults}g), where the hyperpixels exhibits lower condition number than the bandpass.  However, we still note that the hyperpixels perform visibly better than the bandpass filters, e.g. in the 4th column of Figure \ref{fig:imagingresults} where more power appears in the bottom right square.

To robustly quantify any change in condition number in spite of this variation, we perform a $t$-test across 10 different measured areas of the filter array (Figure \ref{fig:imagingresults}h). We find a significant reduction in condition number for hyperpixels compared to bandpass filters (ratio = 3.47x, 95\% CI = 1.28 -- 9.40, $p$ = 0.020).  

\section{Discussion}
We have presented a novel approach to matched filter design for pixel filter arrays that utilizes numerous small `subpixel filter' Fabry-Perot resonators within an image sensor pixel area to create `hyperpixels', whose spectral response matches that of target spectra in a sample.  Though our proposed design operates under a fixed-power constraint---since any given region can still only transmit one spectral band---there should still theoretically be a discernible advantage of this approach when performing spectral unmixing as it better exploits the broad spectral shape of many realistic targets. Indeed, our experimental tests with fabricated hyperpixel filter arrays show, using both measured spectra and imaging, that hyperpixels reduce the condition number of the spectral unmixing matrix at least 2-3 times compared to optimal bandpass filters.  We have thus shown that these designs provide a measurable practical benefit. Considering that our experimental filter arrays are non-optimal (i.e. mismatch with designed structures), our results suggest the potential impact under optimized conditions could be far greater. 

To progress the hyperpixel approach towards real-world application, there are some challenges that must be addressed:  The first challenge is achieving more uniform performance across the fabricated sample.  This could be improved by tighter control of the fabrication process, including exposure and development steps for the grayscale lithography aspect. Using a parallelized exposure scheme such as mask-based photolithography with either a single grayscale mask or a stepper (with dose modulation) may improve uniformity and is also scalable to larger-volume manufacturing.

The second challenge is to modify the design process to better optimize for the the fixed power constraint. Due to the division of pixel area into spectral bands, there is inherently a fixed power constraint: the integrated area under the transmission spectrum of a hyperpixel is equal to that of a bandpass filter. However, previous simulation work has demonstrated that even under this fixed power constraint, a $>$ 5-fold improvement in signal to noise ratio (SNR) under realistic noise conditions can be achieved \cite{taylor-williamsSpectrallyTailoredHyperpixel2022}. If there were no power constraint, we would expect hyperpixels to always perform equal or better than bandpass, with the limiting case being that the two are equivalent (e.g. for bandpass spectral reflectance targets).  However, because of the fixed power constraint of our design there are scenarios in which bandpass filters perform better \cite{taylor-williamsSpectrallyTailoredHyperpixel2022}. Conversely, there are also many scenarios where the hyperpixels, even under the total power constraint, outperform optimal bandpass filters and these will vary depending on the target. 

Our design process for hyperpixels relies on the same assumptions used in designing matched filters, which a fixed total power constraint.  Though this offers simplicity and clearly provides a measurable benefit (as demonstrated), alternative approaches such as water-pouring algorithm or a generic full-spectrum optimization may further maximize condition number \cite{goldsmithCapacityFadingChannels1997}.

Linked to this is the third challenge: improving power transmittance. This could be achieved by using high-index all-dielectric metasurfaces to provide the tailored spectral response, though scaling such approaches to manufacturability scale may be challenging.  However, approaches such as nanoimprint lithography make this option increasingly feasible for production. Spectral responses could also be achieved through multilayer reflectors, as used for multivariate optical elements, although this may require numerous additional fabrication steps to create arrays \cite{yiFabricationMultispectralImaging2011, hooperOptimizingAchromaticityMetalenses2022, wangIntegratedOpticalFilter2006, soyemiDesignTestingMultivariate2001}.

In terms of applications, we note that hyperpixels do not provide benefit in all scenarios. For example if the expected spectral response of the targets resemble bandpass filters, e.g. detecting fluorescence from multiple fluorophores, the optimal hyperpixel design will be similar to the bandpass filter design.  However, the hyperpixel approach offers advantages in specific applications where the target spectra span large spectral bandwidths.  Because of the high customisability, we anticipate hyperpixels could find use for low-volume applications where real-time performance is needed and unmixing is typically noisy. By using techniques such as photolithography or nanoimprint lithography, one could make standalone filter arrays in moderate quantities to be bonded to pre-packaged monochrome CMOS sensors rather than fabricating directly on sensors during manufacturing.  Potential applications may include machine vision where high volumes of items with \emph{a priori} known spectra are encountered, or biomedical spectral imaging where common analyte spectra such as oxygenated and deoxygenated hemoglobin are imaged across a wide range of applications.

\section{Conclusion}
Overall, we conclude that hyperpixel filter arrays can provide a measurable performance benefit over optimal band-pass filter arrays despite the aforementioned fabrication limitations.  They are therefore a highly promising technology for bespoke spectral imaging systems.

\begin{backmatter}
\bmsection{Funding}
The authors acknowledge support from a UKRI Future Leaders Fellowship (MR/T041951/1, GSDG), Cancer Research UK (A9545/A29580, SEB) and EPSRC (EP/R003599/1, SEB), and a NanoPrime grant from the Nottingham Nano- and Micro-scale Research Centre (EP/R025282/1).

\bmsection{Acknowledgments}
The authors thanks the Nottingham Nano- and Micro-scale Research Centre (nmRC) for fabricating all samples.

\bmsection{Disclosures}
The authors declare no conflicts of interest.

\bmsection{Data Availability Statement}
Data in this study available at \cite{datadoi}.

\end{backmatter}


\bibliography{main_v2}

\end{document}